\documentclass[iop]{emulateapj}

\usepackage{amsmath,amssymb,gensymb,times,graphics,morefloats,color}
\usepackage{afterpage, bm}
\usepackage{natbib,hyperref}
\def\kepler{\emph{Kepler}}

\def\mstar{M_*}
\def\rstar{R_*}
\def\msun{M_{\odot}}

\def\rsun{R_{\odot}}
\def\teff{T_\mathrm{eff}}

\def\mk{\mathrm{M}_\mathrm{K}}

\shorttitle{Stellar rotation and the planetary habitable zone}
\shortauthors{Newton et al.}

\begin{document}

\title{The impact of stellar rotation on the detectability of habitable planets around M dwarfs}

\author{Elisabeth R. Newton\altaffilmark{1}, Jonathan Irwin\altaffilmark{1}, David Charbonneau\altaffilmark{1}, Zachory K. Berta-Thompson\altaffilmark{2}, Jason A. Dittmann\altaffilmark{1}}
\altaffiltext{1}{Harvard-Smithsonian Center for Astrophysics, 60 Garden Street, Cambridge, MA 02138, USA}
\altaffiltext{2}{Kavli Institute for Astrophysics and Space Research and Department of Physics, Massachusetts Institute of Technology, Cambridge, MA 02139, USA}

\begin{abstract}

Stellar activity and rotation frustrate the detection of exoplanets through the radial velocity technique. This effect is particularly of concern for M dwarfs, which can remain magnetically active for billions of years. We compile rotation periods for late-type stars and for the M dwarf planet-host sample in order to investigate the rotation periods of older field stars across the main sequence. We show that for stars with masses between $0.25$ and $0.5$ $\msun$ (M4V--M1V), the stellar rotation period typical of field stars coincides with the orbital periods of planets in the habitable zone. This will pose a fundamental challenge to the discovery and characterization of potentially habitable planets around early M dwarfs. Due to the longer rotation periods reached by mid M dwarfs and the shorter orbital period at which the planetary habitable zone is found, stars with masses between $0.1$ and $0.25$ $\msun$ (M6V--M4V) offer better opportunities for the detection of habitable planets via radial velocities.

\end{abstract}

\section{Introduction}

Radial velocities are a powerful tool for the discovery and characterization of exoplanetary systems. However, stellar magnetic activity and rotation affect the measured radial velocities, as surface inhomegeneities cross the stellar surface and change with time. Starspots diminish light received from the limbs of a rotating star, causing changes in spectral line centroid \citep[e.g.][]{Saar1997}. Plages have a similar influence, but with the opposite sign \citep[e.g.][]{Meunier2010}. Strong localized magnetic fields, typically associated with spots, modify convective flows and cause a net blueshift \citep[e.g.][]{Gray2009, Meunier2010}. Disentangling these stellar activity signals from the planetary reflex motion is challenging, and typically requires modeling activity-induced changes in the spectral line profiles \citep[e.g.][]{Queloz2009, Boisse2011, Dumusque2011, Dumusque2012, Rajpaul2015}. This is particularly pertinent for M dwarf stars, which retain high levels of magnetic activity for longer than Sun-like stars. Studies of exoplanets around the nearby M dwarfs Gl 581, Gl 667C, and Gl 191 (Kapteyn's star) have recently highlighted this concern.

Gl 581 is an M3V \citep{Henry1997}\footnote{Included in the RECONS compilation: \url{http://www.recons.org/TOP100.posted.htm}} star. \citet{Bonfils2005} first discovered the hot Neptune Gl 581 b orbiting this star, while \citet{Udry2007} identified a second short-period planet, Gl 581 c. The number of additional planets that orbit this star has been a matter of recent debate. \citet{Udry2007} further identified planet d, and \citet{Mayor2009} planet e. \citet{Vogt2010} found evidence for a long-period planet f, and a potentially habitable planet g. Analyses by other groups quickly cast doubt on planets f \citep{Tuomi2011, Forveille2011} and g \citep{Gregory2011, Baluev2012, Hatzes2013}. \citet{Baluev2012} also raised concerns about planet d. \citet{Robertson2014} looked at the H$\alpha$ line, a magnetic activity indicator. They found support for the existence of the three innermost planets (b, c, and e), but indications that the outer planets (d, g, and f) are the result of stellar activity and rotation.

\citet{Bonfils2013} identified a super-Earth around Gl 667C, with a second found by \citet{Anglada-Escude2012a} and \citet{Delfosse2013}. \citet{Anglada-Escude2013} identified a total of six Keplerian signals at a range of periods. Though long-period signals were seen by \citet{Delfosse2013} and by \citet{Makarov2014}, they were postulated to be related to the $\sim105$ day stellar rotation period. Independent analyses by \citet{Feroz2013} and \citet{Robertson2014a} did not find evidence for planets exterior to Gl 667C b and c.

\citet{Anglada-Escude2014} reported two planets with periods of $48$ and $120$ days around Gl 191.  \citet{Robertson2015} showed that the $48$ day signal is correlated with stellar activity, and suggested a non-planetary origin. They also express concern that the period of the outer planet is close to the $143$ day stellar rotation period.

These examples illustrate that stellar activity signals, which appear at the stellar rotation period and its harmonics, can mimic the effect of a planet at or near these orbital periods. The typical rotation periods of field-age dwarf stars varies by almost an order of magnitude across the lower main sequence, increasing from about $25$ days for early G stars \citep{Barnes2007}, to greater than $100$ days for mid-to-late M dwarfs \citep{Newton2016}. Over the same stellar mass range, the orbital period corresponding to a habitable planet decreases from approximately $365$ days to $10$ days. There thus exists a range of stellar masses where the stellar rotation and planetary habitable-zone periods overlap. This came to our team's attention in pursuit of follow-up observations of K2-3, an early M dwarf with three confirmed transiting super-Earths \citep{Crossfield2015}: we noted that the stellar rotation period of $\sim40$ days is very close to the $44$ day orbital period of K2-3d, which is near the inner edge of the planetary habitable zone.

There are several current and near-future radial velocity surveys committed to finding planets around M dwarf stars, including CARMENES \citep{Quirrenbach2014}, HPF \citep{Mahadevan2012}, and SPIRou \citep{Artigau2014}. It is therefore critical that we understand how stellar rotation may impact the discovery of the potentially habitable planets around these stars. In this Letter, we leverage new rotation periods of mid-to-late M dwarfs in the solar neighborhood from our recent work in \citet{Newton2016} to investigate this problem for stars across the M spectral class.

\section{Stellar rotation periods and masses across the main sequence}

\begin{figure*}
\includegraphics{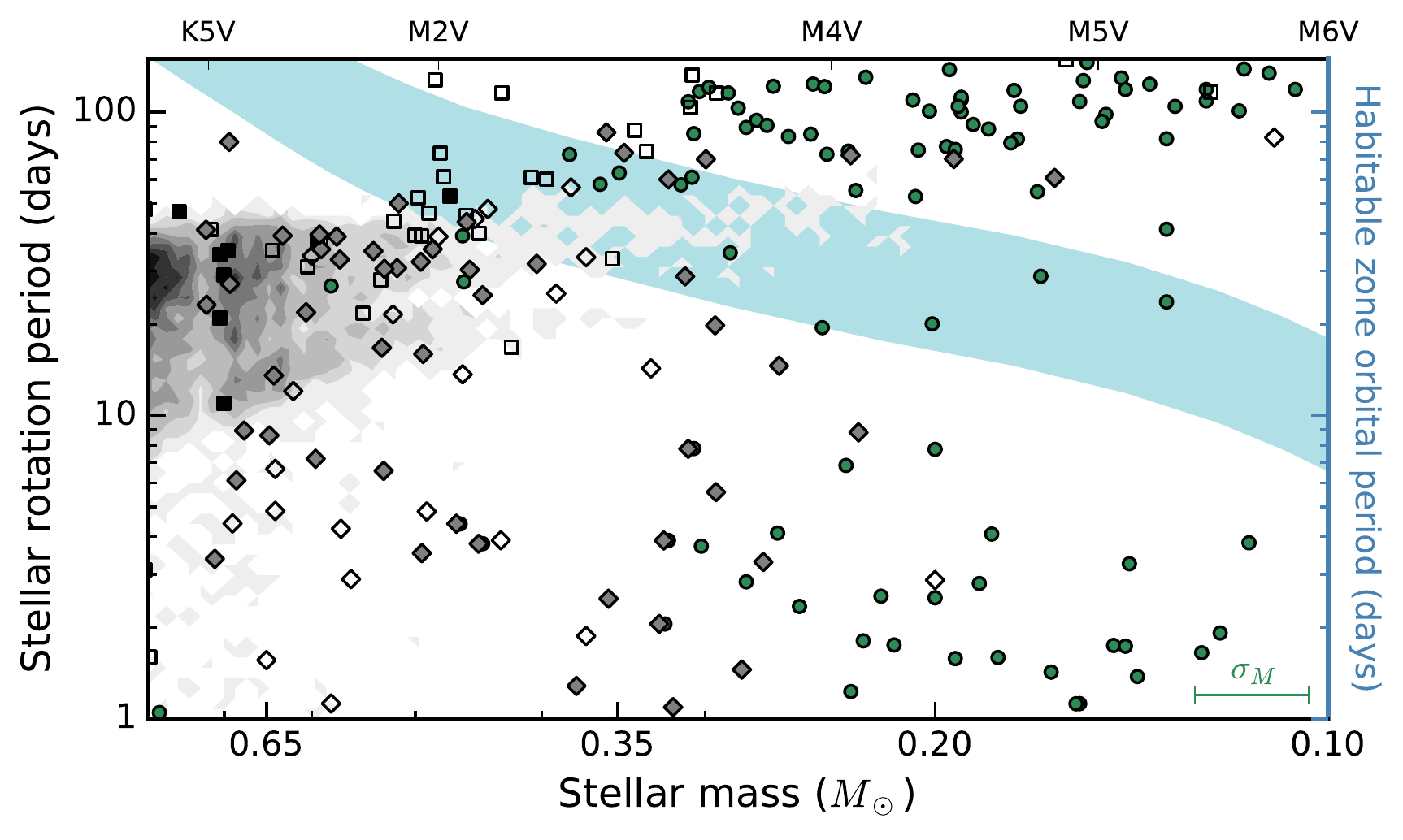}
\caption{Stellar rotation period as a function of stellar mass and its relation to the planetary habitable zone. Gray shading shows the rotation period distribution from \citet{McQuillan2014}, who analyzed data from \kepler. Filled black squares are from \citet{Baliunas1996}, and open black squares are from \citet{SuarezMascareno2015}. Open diamonds are from \citet{Kiraga2007}, and filled diamonds are from \citet{Hartman2011}. Green circles are M dwarf rotators from the \citet{Newton2016} statistical sample. The upper envelope indicates the rotation periods of typical, older field stars. The error bar in the lower right indicates the $10\%$ scatter in the empirical $\mk$--mass relation used to infer stellar mass for the M and and late K dwarfs \citep{Delfosse2000}. The blue shaded region shows the habitable zone as a function of stellar mass from \citet{Kopparapu2013}. For early M dwarfs, the habitable zone period overlaps with the rotation periods of typical stars, both of which are about 40 days. This coincidence complicates the radial velocity detection of habitable planets around early M dwarfs, but not for mid-to-late M dwarfs, or for G and K dwarfs. \label{Fig:hzharmonics}}
\end{figure*}

To assess stellar rotation across the main sequence, we compile literature measurements for the rotation periods of field stars, using data from \citet{Baliunas1996}, \citet{Kiraga2007}, \citet{Hartman2011}, \citet{SuarezMascareno2015}, and \citet{Newton2016}. The latter supplies most of the rotation period measurements for mid and late M dwarfs, and results from our analysis of photometry from the MEarth-North transit survey \citep{Berta2012,Irwin2014}. The rotation periods from \citet{Hartman2011} and \citet{Kiraga2007} also come from broadband optical or red-optical photometry. On the other hand, \citet{Baliunas1996} and \citet{SuarezMascareno2015} measure periodicities in chromospheric activity features. 

We use stellar mass ($\mstar$) as the independent parameter for this study. For G and early K dwarfs, we use $B-V$ colors to estimate effective temperature ($\teff$), and then use isochrones to infer $\mstar$. We use the $B-V$ colors provided by \citet{Baliunas1996} and \citet{SuarezMascareno2015} in their works. We use the empirical relation between $B-V$ color and $\teff$ from \citet{Boyajian2013} to determine $\teff$. We adopt the $2$ Gyr solar-metallicity isochrones from \citet[][]{Baraffe2015}.

For late K and M dwarfs, there are known discrepancies between the theoretical and observed stellar temperatures and masses \citep[e.g.][]{Boyajian2012}. For stars with $B-V>1$, we therefore use the $\mk$--mass relation from \citet{Delfosse2000}, which we have modified to allow extrapolation as described in \citet{Newton2016}. \citet{Delfosse2000} calibrated an empirical relation to determine stellar mass by relating the dynamical masses measured from binary orbits to the stars' absolute CIT K magnitudes ($\mk$). The scatter in the relation is about $10\%$.  We restrict the K and M dwarfs included in this study to those with trigonometric distances. We use NIR magnitudes from 2MASS \citep[, requiring \texttt{qual\_flag} = AAA]{Cutri2003}, or alternatively CIT magnitudes from \citet{Leggett1992}.

We also include the rotation periods measured from \kepler\ photometry by \citet{McQuillan2014}. Most \kepler\ stars do not have parallaxes and they may be subject to interstellar reddening. Therefore, we use the $\teff$ listed in \citet{McQuillan2014}. These $\teff$ are adopted from the \kepler\ Input Catalog \citep[KIC;][]{Brown2011} or, where available, from \citet{Dressing2013}. We then interpolate $\teff$ onto the $2$ Gyr solar-metallicity isochrones from \citet[][]{Baraffe2015} to infer stellar mass. The difference in the way masses are estimated for the \kepler\ sample may result in an offset relative to the masses of K and M dwarfs estimated from $\mk$ \citep[e.g.][]{Newton2015}.

Figure \ref{Fig:hzharmonics} shows stellar rotation period as a function of stellar mass for late-type stars in the field. Given the uncertainties in the stellar parameters for the \kepler\ sample, we treat these as an ensemble and plot the distribution instead of the individual measurements. The sensitivity of the \citet{McQuillan2014} analysis may be limited by \kepler's quarterly flux offsets; the other surveys included in this compilation detected periods out to $100$ days or longer.

Figure \ref{Fig:hzharmonics} includes stars of a range of ages. As single stars age, they lose angular momentum and their rotation periods increase. Angular momentum loss is mass dependent, with lower-mass stars taking longer to spin down, but eventually reaching longer rotation periods. The long-period envelope of the rotation period distribution is thought to represent an old field population, and the fast rotators at a given mass will eventually evolve toward these long periods\footnote{There are handful of early M dwarfs with rotation periods of around $100$ days, which stand out as lying above this envelope. Though intriguing, discussion is beyond the scope of this paper.}. The envelope comprises stars of similar age to the Sun and older, which was established by \citet{Skumanich1972} and \citet{Barnes2003} for Sun-like stars. We quantified this for M dwarfs by using galactic kinematics to estimate that mid M dwarfs with rotation periods around $100$ days are on average $5$ Gyr old \citep{Newton2016}.

\section{Stellar rotation and the habitable zone}\label{Sec:planets}

When radial velocity variations induced by stellar rotation and activity coincide with the planetary orbital period, it can be difficult to identify the planetary signal. This stellar signal can appear not just at the rotation period itself, but at higher-order harmonics as well. \citet{Boisse2011} tested cases where the photosphere is dominated by one or two spots and found that power appears primarily at one-half and one-third the stellar rotation period \citep[see also][]{Aigrain2012}. Certain configurations resulted in a low-amplitude signal at one-fourth the rotation period as well. Though spot patterns are likely more complex, both \citet{Boisse2011,Boisse2012} and \citet{Aigrain2012} had success modeling observed radial velocity data.  With typical observational sampling, power from these stellar signals leak into nearby frequencies \citep[see, e.g., Fig. 15 in][]{Boisse2011}. Thus, stars with rotation periods that are the same as, or two to three times longer than, the habitable zone are not optimal targets for radial velocity searches that aim to discover potentially habitable planets.

\begin{figure}
\includegraphics[width=3.5in]{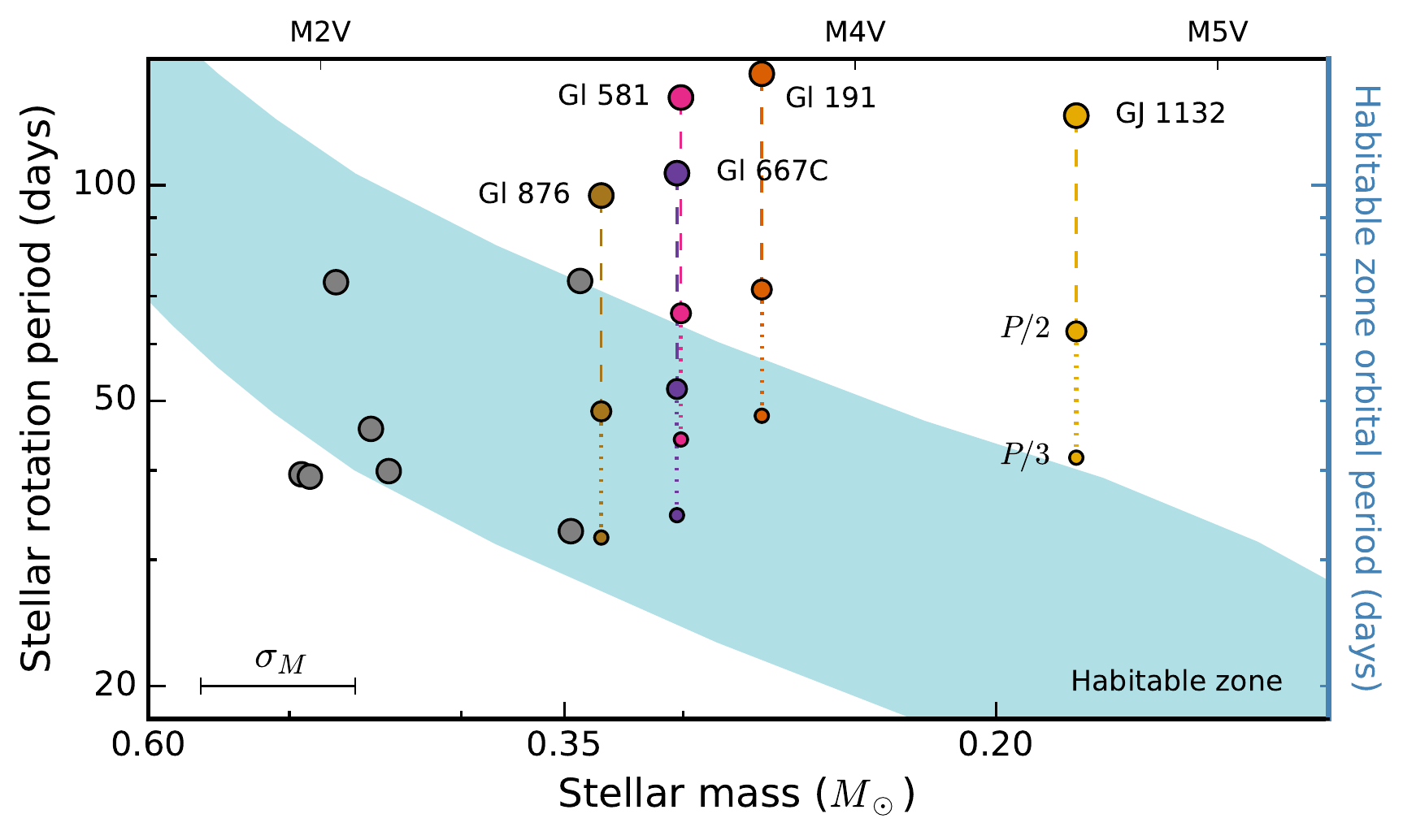}
\caption{Stellar rotation period vs. stellar mass and its relation to the planetary habitable zone for M dwarfs planet hosts. The error bar in the lower left indicates the $10\%$ scatter in the empirical $\mk$--mass relation used to infer stellar mass \citep{Delfosse2000}. The blue shaded region shows the habitable zone as a function of stellar mass from \citet{Kopparapu2013}. For early M dwarfs, the habitable-zone period overlaps with the rotation periods of typical stars, both of which are about 40 days. GJ 1214 has been excluded from this plot due to the range of published rotation periods (see Table \ref{Tab:planets}). \label{Fig:planets}}
\end{figure}

\begin{figure}
\includegraphics[width=3.5in]{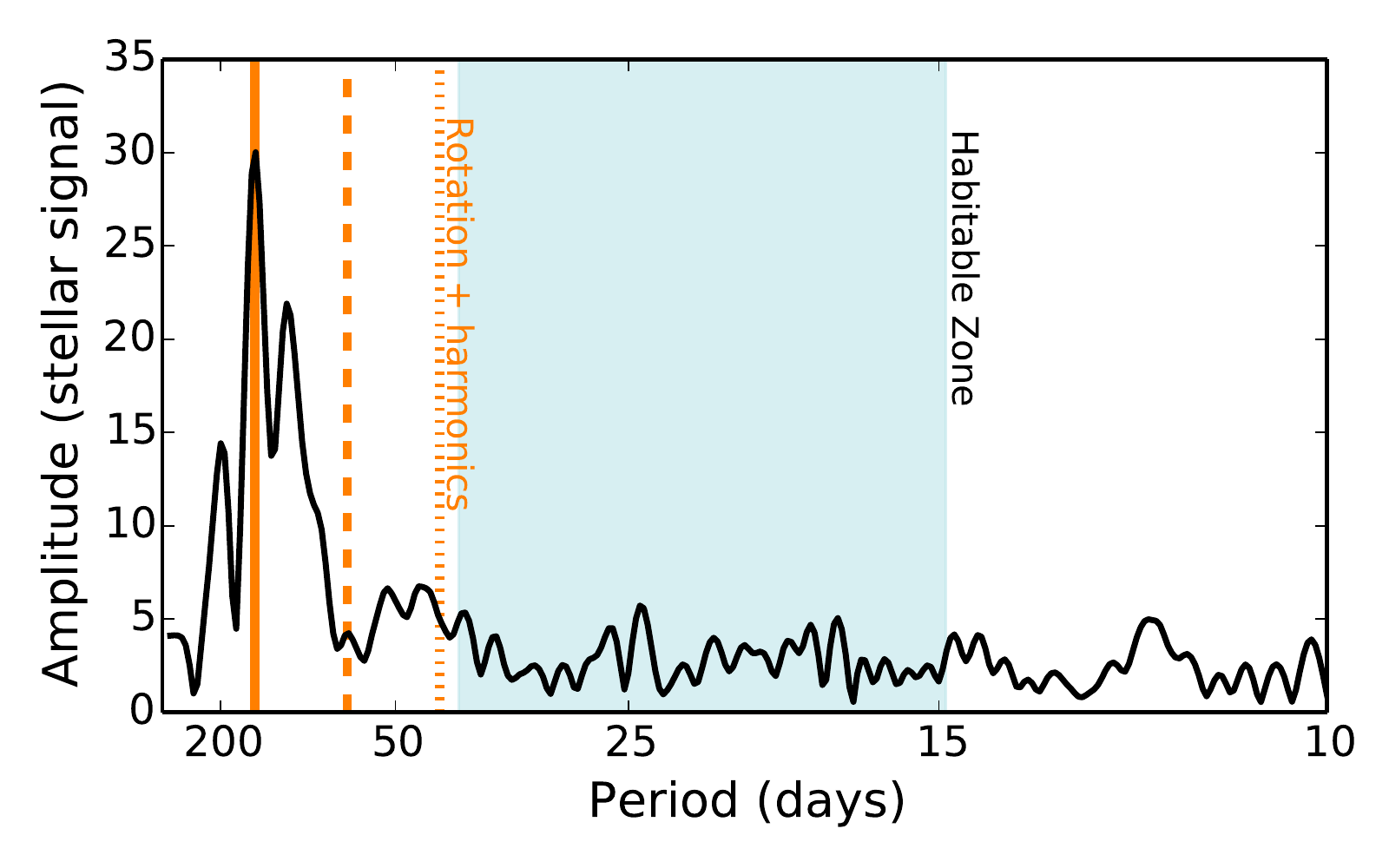}
\caption{Periodogram of the MEarth light curve for GJ 1132 \citep{Berta-Thompson2015}. The highest peak, indicated by the solid vertical line, is the adopted rotation period. The second and third harmonics are indicated by dashed and dotted lines, respectively; the radial velocity signal from the starspots associated with its photometric variability is expected to have power at these frequencies. The shaded region indicates the planetary habitable zone. The frequency space within which one would find a habitable planet is not expected to be strongly contaminated by stellar activity.\label{Fig:gj1132}}
\end{figure}

The habitable zone from \citet[][using values from the erratum]{Kopparapu2013} is indicated in Figures \ref{Fig:hzharmonics}-\ref{Fig:gj1132}. We adopt the moist greenhouse limit as the inner edge of the habitable zone and the maximum greenhouse as the outer edge of the habitable zone. We use the \citet{Baraffe2015} solar-metallicity $2$ Gyr isochrone to calculate $\teff$ and luminosity ($L$) at a given stellar mass. We note that the choice of stellar models as well as parameters beyond stellar mass, in particular stellar age and metallicity, affect $\teff$ and $L$, and therefore also the precise location of the calculated habitable zone.

For $0.35$ $\msun<\mstar<0.5$ $\msun$, Figure \ref{Fig:hzharmonics} shows that the orbital period of habitable-zone planets is similar to the rotation period of the typical $5$ Gyr star. Power can also appear at multiples of the stellar rotation period, so activity signatures can pose problems for habitable-zone planets even with slowly rotating early M dwarfs. This means that stellar rotation can be a confounding factor for $0.25\msun<\mstar<0.5\msun$. This corresponds to spectral types between M1V and M4V.

We illustrate this idea by considering known M dwarf planet hosts (Fig. \ref{Fig:planets}). Our planet-host sample is drawn from the \texttt{exoplanet.eu} database \citep{Schneider2011}. We select those stars with $\teff<4200$ K, $\mstar<0.7$ $\msun$, $\rstar<0.7$ $\rsun$, and a measured trigonometric parallax. We use the $\mk$--mass relation from \citet{Delfosse2000} to estimate stellar masses. We use parallaxes from \citet[][from \textit{Hipparcos}]{VanLeeuwen2007}, and supplement with additional sources as necessary. We adopt NIR magnitudes from 2MASS \citep[, requiring \texttt{qual\_flag} = AAA]{Cutri2003}. Where 2MASS magnitudes are not available, we use CIT magnitudes from \citet{Leggett1992}. We then use the \citet{Delfosse2000} relation to estimate stellar masses and the \citet{Boyajian2012} single-star relation to estimate stellar radii. The \citet{Delfosse2000} relation is based on CIT magnitudes, so we convert 2MASS magnitudes to the CIT system using the relations from \citet{Cutri2006}. Stellar rotation periods are compiled from the literature, though many are from \citet{SuarezMascareno2015}. We do not consider estimates from rotational broadening of spectral lines, which are subject to uncertainty in the inclination angle. These data are listed in Table \ref{Tab:planets}.

Figure \ref{Fig:planets} shows the rotation periods of these planet hosts. Gl 581, in particular, highlights the challenge early M dwarfs pose for radial velocity surveys seeking to find habitable planets. Though this star has a very long rotation period ($130$ days), the harmonics dip into the habitable zone. This resulted in the controversy over the potentially habitable planet Gl 581 d ($P_\mathrm{d}=66$ days, at about $P_\mathrm{rot}/2$), as well as g ($P_\mathrm{g}=36$ days, between $P_\mathrm{rot}/3$ and $P_\mathrm{rot}/4$).

\begin{deluxetable*}{l l l l l l l l}
\tablecolumns{8}
\tablecaption{Stellar parameters for M dwarf planet hosts with trigonometric parallaxes \label{Tab:planets}
}
\tablehead{
	\colhead{Star} &
	\colhead{Distance} &
	\colhead{Ref.\tablenotemark{a}} &
	\colhead{$K_S$} &
	\colhead{Mass} &
	\colhead{Radius} &
	\colhead{$P_\mathrm{rot}$} &
	\colhead{Ref.\tablenotemark{b}} \\
	&
	\colhead{(pc)} &
	&
	\colhead{(2MASS)} &
	\colhead{($\msun$)} &
	\colhead{($\rsun$)} &
	\colhead{(days)} &
	 }
\startdata
\sidehead{Published rotation period}
\hline \\
Gl 176 & 9.27 & vL07 & 5.607 & 0.49 & 0.47 & 39.5 & R15a \\ 
Gl 191 & 3.91 & vL07 & 5.049 & 0.27 & 0.28 & 143.0 & R15b \\ 
Gl 433 & 8.88 & vL07 & 5.623 & 0.47 & 0.45 & 73.2 & SM15 \\ 
Gl 436 & 10.14 & vL07 & 6.073 & 0.44 & 0.42 & 39.9 & SM15 \\ 
Gl 581 & 6.21 & vL07 & 5.837 & 0.30 & 0.30 & 132.5 & SM15 \\ 
Gl 667C & 6.84 & vL07 & 6.036 & 0.30 & 0.30 & 103.9 & SM15 \\ 
Gl 674 & 4.54 & vL07 & 4.855 & 0.35 & 0.34 & 32.9 & SM15 \\ 
Gl 832 & 4.95 & vL07 & 4.461 & 0.45 & 0.43 & 45.7 & SM15 \\ 
Gl 849 & 9.1 & vL07 & 5.594 & 0.49 & 0.46 & 39.2 & SM15 \\ 
Gl 876 & 4.69 & vL07 & 5.010 & 0.33 & 0.33 & 96.7 & R05 \\ 
GJ 1132 & 12.04 & J05 & 8.322 & 0.18 & 0.21 & 125.0 & BT15 \\ 
GJ 1214 & 14.55 & AE13 & 8.782 & 0.18 & 0.21 & $\cdots$ & (See note) \\ 
HIP 57050 & 11.1 & vL07 & 6.822 & 0.34 & 0.34 & 73.5 & H11 \\ 
\sidehead{No rotation period available}
\hline \\
Gl 15A & 3.59 & vL07 & 4.011 & 0.40 & 0.39 & $\cdots$ & $\cdots$ \\ 
Gl 27.1 & 23.99 & vL07 & 7.394 & 0.55 & 0.52 & $\cdots$ & $\cdots$ \\ 
Gl 163 & 14.99 & vL07 & 7.135 & 0.40 & 0.38 & $\cdots$ & $\cdots$ \\ 
Gl 179 & 12.29 & vL07 & 6.942 & 0.36 & 0.35 & $\cdots$ & $\cdots$ \\ 
Gl 180 & 12.12 & vL07 & 6.598 & 0.41 & 0.40 & $\cdots$ & $\cdots$ \\ 
Gl 229 & 5.75 & vL07 & 4.131 & 0.58 & 0.55 & $\cdots$ & $\cdots$ \\ 
Gl 317 & 15.3 & AE12 & 7.028 & 0.43 & 0.41 & $\cdots$ & $\cdots$ \\ 
Gl 328 & 19.79 & vL07 & 6.352 & 0.68 & 0.65 & $\cdots$ & $\cdots$ \\ 
Gl 422 & 12.67 & vL07 & 7.035 & 0.35 & 0.35 & $\cdots$ & $\cdots$ \\ 
Gl 649 & 10.34 & vL07 & 5.624 & 0.54 & 0.51 & $\cdots$ & $\cdots$ \\ 
Gl 682 & 5.08 & vL07 & 5.606 & 0.27 & 0.28 & $\cdots$ & $\cdots$ \\ 
Gl 687 & 4.53 & vL07 & 4.501 & 0.41 & 0.39 & $\cdots$ & $\cdots$ \\ 
GJ 3341 & 23.2 & R10 & 7.733 & 0.47 & 0.44 & $\cdots$ & $\cdots$ \\ 
GJ 3634 & 19.8 & R10 & 7.470 & 0.45 & 0.43 & $\cdots$ & $\cdots$ \\ 
HIP 12961 & 23.01 & vL07 & 6.736 & 0.67 & 0.64 & $\cdots$ & $\cdots$ \\ 
HIP 70849 & 23.57 & vL07 & 6.790 & 0.67 & 0.64 & $\cdots$ & $\cdots$ \\ 
HIP 79431 & 14.4 & vL07 & 6.589 & 0.49 & 0.46 & $\cdots$ & $\cdots$ \\ 
Wolf 1061 & 4.29 & vL07 & 5.075 & 0.30 & 0.30 & $\cdots$ & $\cdots$ 
\enddata
\tablenotetext{a}{References for trigonometric distances:~
  J05 = \citet{Jao2005};
  vL07 = \citet{VanLeeuwen2007};
  R10 = \citet{Riedel2010};
  AE12 = \citet{Anglada-Escude2012};
  AE13 = \citet{Anglada-Escude2013a}
  }
\tablenotetext{b}{References for stellar rotation periods:~
  R05 = \citet{Rivera2005};
  H11 = \citet{Hartman2011};
  R15a = \citet{Robertson2015a};
  R15b = \citet{Robertson2015};
  BT15 = \citet{Berta-Thompson2015};
  SM15 = \citet{SuarezMascareno2015}
  }
\tablecomments{
GJ 1214 displays photometric modulation, but has not yielded a secure rotation period detection. \citet{Berta2011} tentatively derived a rotation period of $52$ days from MEarth photometry and noted that the long-period signal in one season of data indicates that the period may be $80-100$ days. Our recent re-analysis in \citet{Newton2016} did not yield a firm period detection. Using independent data sets, \citet{Narita2013} and \citet{Nascimbeni2015} found periods of $44$ days and approximately $80$ days, respectively.}

\end{deluxetable*}

GJ 1132 ($\mstar=0.18\msun$) demonstrates that this is not an issue for many of the lowest-mass M dwarfs. Like many stars of this mass, GJ 1132 has a rotation period \citep[$P_\mathrm{rot}=125$ days][]{Berta-Thompson2015} comparable to that of Gl 581. We show the periodogram of this data set in Figure \ref{Fig:gj1132}; there is no significant power at periods shorter than around $60$ days. In the photometry, we do not see power at the second and third harmonics, which is not surprising given the near-sinusoidal modulation seen in \citet{Berta-Thompson2015}. However, power could still appear at these frequencies in the radial velocities. However, its lower-mass results in shorter periods for potentially habitable planets, and the third harmonic of the stellar rotation period still lies outside the outer edge of the habitable zone (although only barely so).

\section{M dwarfs with photometric monitoring as targets for radial velocity surveys}\label{Sec:noperiod}

We have demonstrated that long-period, low-mass rotators provide good targets for radial velocity surveys searching for habitable planets. We have provided a catalog of stars with detected photometric rotation periods in \citet{Newton2016}, which will be useful in selecting such targets. We recommend selecting those targets with long photometric rotation periods (grade A or B rotators, $P\gtrsim100$ days). 

Also of interest to the community will be those stars that may be photometrically quiet. \citet{Newton2016} catalog stars for which we did not detect periodic photometric modulation in the MEarth data. For some of these objects, the light curves were noisy or sparse, which could have prohibited period detection. Further monitoring may indicate whether these stars are good targets for radial velocity surveys. The statistical sample we identified includes the objects more likely to be of immediate interest.

Our statistical sample was the subset of stars for which the number of observations ($n_\mathrm{visits}$) and median error ($\sigma$) were such that we had good sensitivity to a range of rotation periods  ($n_\mathrm{visits}>1200$, $\sigma<0.005$). Within this sample, we did not detect a periodic signal for $52\%$. A Kolmogrov-Smirnov test indicates that the distribution of space velocities for the non-detections is statistically indistinguishable from that of the long-period ($P_\mathrm{rot}>70$ days) sample. We infer that, on average, these stars have ages similar to the long-period sample: we generally expect them to be members of the old thin disk and to be slowly rotating. The stars in the statistical sample with no period detections are therefore also promising targets for radial velocity surveys.

\section{Discussion}

We have demonstrated that the habitable zone of early M dwarfs coincides with the stellar rotation period typical for older field stars (approximately $5$ Gyr). Because activity signals are also expected at two and three times the frequency of the stellar rotation, the stars for which this is an issue are those with $0.25\msun<\mstar<0.5\msun$ (roughly M1V--M4V). This means that stellar activity will be an unavoidable contaminant in radial velocity surveys that are searching for habitable planets around early M dwarfs. Rather than being the exception, we expect that challenges faced in understanding the Gl 581 system are typical for stars with its mass.

The lifetime of stellar activity and of rapid rotation is another important consideration, since precise radial velocity measurements are challenging for stars with strong activity signature and those with rotationally broadened spectral features. \citet{West2008} found that the activity lifetime rises sharply around a spectral type of M4V. Stellar magnetic activity diminishes by $1$ Gyr for M0--M3 dwarfs, but may last for $7-8$ Gyr for M6--M8 dwarfs. It is approximately $4$ Gyr for M4--M5 dwarfs. In what is presumably a related phenomenon, lower-mass stars also retain rapid rotation rates for longer than do higher mass stars: few early M dwarfs are found to be rapidly rotating \citep{McQuillan2013}, while nearly all very late M (and L) dwarfs have detectable rotational broadening \citet{Mohanty2003}. In \citet{Newton2016}, we found an abundance of slowly rotating mid M dwarfs (M4V--M6V), and estimated that a rapid loss of angular momentum occurs  between $2$ and $5$ Gyr for mid M dwarfs.  These considerations disfavor the latest-type M dwarfs, since strong activity signatures and rapid rotation are expected to persist for a substantial part of the stars' lifetime. 

Finally, we consider the prospects for detailed characterization. A key motivation for turning to M dwarfs is that their small stellar size enhances the signals of orbiting planets. However, M dwarfs span nearly a decade in stellar mass and radius. This means that the transit and radial velocity signatures of a planet orbiting a mid-to-late M dwarf are significantly larger than if an analogous planet orbited an early M dwarf. For example, the transit depth of the Earth around the Sun is  $0.0084\%$; for the same size planet orbiting an $\rstar=0.5\rsun$ star (M1V), the transit depth is $0.034\%$, while for an $\rstar=0.15\rsun$ star (M5V), it is $0.37\%$. Similarly, an Earth-mass planet at the inner edge of the habitable zone induces a radial velocity semi-amplitude of $0.25$ m/s if it orbits an $\mstar=0.55\msun$ star (M1V), compared to $0.98$ m/s if it orbits an $\mstar=0.15\msun$ star (M5V).  

A promising avenue to look for biosignatures on habitable exoplanets is the direct detection of the planetary Doppler shift. \citet{Snellen2013} recently explored the prospects for detecting oxygen by looking at exoplanetary telluric features. \citet{Rodler2014} undertook a detailed simulation of the feasibility of this detection if using a high-resolution spectrograph on an Extremely Large Telescope. Their work indicates that the detection of oxygen in the atmosphere of a habitable, Earth-sized planet is a significant time investment that will only be feasible for nearby M dwarfs with $\mstar<0.25\msun$. This is due to the relative size of the star and the planet, and the orbital periods of planets in the habitable zone. Such a measurement would be within reach only for the nearest and brightest mid-to-late M dwarfs.

We have considered three factors critical to understanding the prospects of detecting and characterizing habitable planets around M dwarfs: 
\begin{enumerate} 
\item{The coincidence of the habitable-zone orbital periods and stellar rotation periods.}
\item{The lifetime of activity and rapid rotation.}
\item{The accessibility of detailed atmospheric characterization.}
\end{enumerate}
We have demonstrated that the confluence of these factors make mid M dwarfs ideal targets for radial velocity surveys that have the goal of discovering habitable planets whose atmospheres can be studied with the \emph{James Webb Space Telescope} and the ELTs. Many of these targets will be discovered in the coming years by radial velocity instruments and transit surveys (the latter of which are  likely to receive radial velocity follow-up). Early M dwarfs are tempting targets for such surveys: their relative brightness makes it easier to achieve higher signal-to-noise observations and more precise radial velocity measurements. However, stellar activity signals will be present at the same periods at which the habitable planets are found. Recent debate surrounding the candidate exoplanets orbiting Gl 581, Gl 667C, and Gl 191 have already demonstrated the challenge of disentangling planetary reflex motion from stellar activity signals.

We therefore suggest that radial velocity surveys closely consider the mid M dwarfs, with $0.1\msun<\mstar<0.25\msun$ (M4V--M6V). Recent work by the MEarth team provides a useful starting place from which to build such a sample in the northern hemisphere. \citet{Nutzman2008} selected M dwarfs from the \citet{Lepine2005a} northern proper motion catalog with estimated stellar radii $\rstar<0.33\rsun$ and trigonometric or photometric distances that placed them within $33$ pc. We have since obtained low-resolution near-infrared and optical spectroscopy for a quarter of the sample, and astrometric and photometric analysis of the entire MEarth-North data set. The near-infrared spectroscopic survey conducted by \citet{Terrien2015} also includes substantial overlap with the MEarth-North sample, as do the optical spectroscopic surveys from \citet{Lepine2013}, \citet{Gaidos2014}, and \citet{Alonso-Floriano2015}. Detailed characterization from the MEarth team includes:
\begin{itemize}
\item{Kinematic radial velocities \citep{Newton2014}.}
\item{Spectroscopically-estimated stellar parameters \citep{Newton2015}.}
\item{H$\alpha$ measurements \citep[][and Newton et al.\ in prep]{West2015}.}
\item{Trigonometric parallaxes \citep{Dittmann2014}.}
\item{Optical colors and estimated metallicities \citep{Dittmann2015}.}
\item{Photometric rotation periods and non-detections \citep{Newton2016}.}
\end{itemize}

\acknowledgments The MEarth project acknowledges funding from the National Science Foundation under grants AST-0807690, AST- 1109468, and AST-1004488 (Alan T. Waterman Award) and the David and Lucile Packard Foundation Fellowship for Science and Engineering. This publication was made possible through the support of a grant from the John Templeton Foundation. The opinions expressed here are those of the authors and do not necessarily reflect the views of the John Templeton Foundation. E.R.N. was supported by the NSF Graduate Research Fellowship, and Z.K.B.-T. by the MIT Torres Fellowship for Exoplanet Research. This research has made use of data products from the Two Micron All Sky Survey, which is a joint project of the University of Massachusetts and the Infrared Processing and Analysis Center / California Institute of Technology, funded by NASA and the NSF; NASA Astrophysics Data System (ADS); and the SIMBAD database and VizieR catalog access tool, at CDS, Strasbourg, France.

\clearpage


\begin{thebibliography}{}
\expandafter\ifx\csname natexlab\endcsname\relax\def\natexlab#1{#1}\fi

\bibitem[{Aigrain {et~al.}(2012)Aigrain, Pont, \& Zucker}]{Aigrain2012}
Aigrain, S., Pont, F., \& Zucker, S. 2012, Monthly Notices of the Royal
  Astronomical Society, 419, 3147

\bibitem[{Alonso-Floriano {et~al.}(2015)Alonso-Floriano, Morales, Caballero,
  Montes, Klutsch, Mundt, Cort{\'{e}}s-Contreras, Ribas, Reiners, Amado,
  Quirrenbach, \& Jeffers}]{Alonso-Floriano2015}
Alonso-Floriano, F.~J., Morales, J.~C., Caballero, J.~A., {et~al.} 2015,
  Astronomy {\&} Astrophysics, 577, A128

\bibitem[{Anglada-Escud{\'{e}}
  {et~al.}(2012{\natexlab{a}})Anglada-Escud{\'{e}}, Boss, Weinberger, Thompson,
  Butler, Vogt, \& Rivera}]{Anglada-Escude2012}
Anglada-Escud{\'{e}}, G., Boss, A.~P., Weinberger, A.~J., {et~al.}
  2012{\natexlab{a}}, The Astrophysical Journal, 746, 37

\bibitem[{Anglada-Escud{\'{e}}
  {et~al.}(2013{\natexlab{a}})Anglada-Escud{\'{e}}, Rojas-Ayala, Boss,
  Weinberger, \& Lloyd}]{Anglada-Escude2013a}
Anglada-Escud{\'{e}}, G., Rojas-Ayala, B., Boss, A.~P., Weinberger, A.~J., \&
  Lloyd, J.~P. 2013{\natexlab{a}}, Astronomy {\&} Astrophysics, 551, A48

\bibitem[{Anglada-Escud{\'{e}}
  {et~al.}(2012{\natexlab{b}})Anglada-Escud{\'{e}}, Arriagada, Vogt, Rivera,
  Butler, Crane, Shectman, Thompson, Minniti, Haghighipour, Carter, Tinney,
  Wittenmyer, Bailey, O'Toole, Jones, \& Jenkins}]{Anglada-Escude2012a}
Anglada-Escud{\'{e}}, G., Arriagada, P., Vogt, S.~S., {et~al.}
  2012{\natexlab{b}}, The Astrophysical Journal, 751, L16

\bibitem[{Anglada-Escud{\'{e}}
  {et~al.}(2013{\natexlab{b}})Anglada-Escud{\'{e}}, Tuomi, Gerlach, Barnes,
  Heller, Jenkins, Wende, Vogt, {Paul Butler}, Reiners, \&
  Jones}]{Anglada-Escude2013}
Anglada-Escud{\'{e}}, G., Tuomi, M., Gerlach, E., {et~al.} 2013{\natexlab{b}},
  Astronomy {\&} Astrophysics, 556, A126

\bibitem[{Anglada-Escude {et~al.}(2014)Anglada-Escude, Arriagada, Tuomi,
  Zechmeister, Jenkins, Ofir, Dreizler, Gerlach, Marvin, Reiners, Jeffers,
  Butler, Vogt, Amado, Rodriguez-Lopez, Berdinas, Morin, Crane, Shectman,
  Thompson, Diaz, Rivera, Sarmiento, \& Jones}]{Anglada-Escude2014}
Anglada-Escude, G., Arriagada, P., Tuomi, M., {et~al.} 2014, Monthly Notices of
  the Royal Astronomical Society: Letters, 443, L89

\bibitem[{Artigau {et~al.}(2014)Artigau, Kouach, Donati, Doyon, Delfosse,
  Baratchart, Lacombe, Moutou, Rabou, Par{\`{e}}s, Micheau, Thibault, Reshetov,
  Dubois, Hernandez, Vall{\'{e}}e, Wang, Dolon, Pepe, Bouchy, Striebig,
  H{\'{e}}nault, Loop, Saddlemyer, Barrick, Vermeulen, Dupieux, H{\'{e}}brard,
  Boisse, Martioli, Alencar, do~Nascimento, \& Figueira}]{Artigau2014}
Artigau, {\'{E}}., Kouach, D., Donati, J.-F., {et~al.} 2014, in Proceedings of
  the SPIE, ed. S.~K. Ramsay, I.~S. McLean, \& H.~Takami, Vol. 9147, 914715

\bibitem[{Baliunas {et~al.}(1996)Baliunas, Sokoloff, \& Soon}]{Baliunas1996}
Baliunas, S., Sokoloff, D., \& Soon, W. 1996, The Astrophysical Journal, 457,
  L99

\bibitem[{Baluev(2012)}]{Baluev2012}
Baluev, R.~V. 2012, Monthly Notices of the Royal Astronomical Society, 429,
  2052

\bibitem[{Baraffe {et~al.}(2015)Baraffe, Homeier, Allard, \&
  Chabrier}]{Baraffe2015}
Baraffe, I., Homeier, D., Allard, F., \& Chabrier, G. 2015, Astronomy {\&}
  Astrophysics, 577, A42

\bibitem[{Barnes(2003)}]{Barnes2003}
Barnes, S.~A. 2003, The Astrophysical Journal, 586, 464

\bibitem[{Barnes(2007)}]{Barnes2007}
---. 2007, The Astrophysical Journal, 669, 1167

\bibitem[{Berta {et~al.}(2011)Berta, Charbonneau, Bean, Irwin, Burke,
  D{\'{e}}sert, Nutzman, \& Falco}]{Berta2011}
Berta, Z.~K., Charbonneau, D., Bean, J., {et~al.} 2011, The Astrophysical
  Journal, 736, 12

\bibitem[{Berta {et~al.}(2012)Berta, Irwin, Charbonneau, Burke, \&
  Falco}]{Berta2012}
Berta, Z.~K., Irwin, J., Charbonneau, D., Burke, C.~J., \& Falco, E.~E. 2012,
  The Astronomical Journal, 144, 145

\bibitem[{Berta-Thompson {et~al.}(2015)Berta-Thompson, Irwin, Charbonneau,
  Newton, Dittmann, Astudillo-Defru, Bonfils, Gillon, Jehin, Stark, Stalder,
  Bouchy, Delfosse, Forveille, Lovis, Mayor, Neves, Pepe, Santos, Udry, \&
  W{\"{u}}nsche}]{Berta-Thompson2015}
Berta-Thompson, Z.~K., Irwin, J., Charbonneau, D., {et~al.} 2015, Nature, 527,
  204

\bibitem[{Boisse {et~al.}(2012)Boisse, Bonfils, \& Santos}]{Boisse2012}
Boisse, I., Bonfils, X., \& Santos, N.~C. 2012, Astronomy {\&} Astrophysics,
  545, A109

\bibitem[{Boisse {et~al.}(2011)Boisse, Bouchy, H{\'{e}}brard, Bonfils, Santos,
  \& Vauclair}]{Boisse2011}
Boisse, I., Bouchy, F., H{\'{e}}brard, G., {et~al.} 2011, Astronomy {\&}
  Astrophysics, 528, A4

\bibitem[{Bonfils {et~al.}(2005)Bonfils, Forveille, Delfosse, Udry, Mayor,
  Perrier, Bouchy, Pepe, Queloz, \& Bertaux}]{Bonfils2005}
Bonfils, X., Forveille, T., Delfosse, X., {et~al.} 2005, Astronomy and
  Astrophysics, 443, L15

\bibitem[{Bonfils {et~al.}(2013)Bonfils, Delfosse, Udry, Forveille, Mayor,
  Perrier, Bouchy, Gillon, Lovis, Pepe, Queloz, Santos, S{\'{e}}gransan, \&
  Bertaux}]{Bonfils2013}
Bonfils, X., Delfosse, X., Udry, S., {et~al.} 2013, Astronomy {\&}
  Astrophysics, 549, A109

\bibitem[{Boyajian {et~al.}(2012)Boyajian, von Braun, van Belle, McAlister, ten
  Brummelaar, Kane, Muirhead, Jones, White, Schaefer, Ciardi, Henry,
  L{\'{o}}pez-Morales, Ridgway, Gies, Jao, Rojas-Ayala, Parks, Sturmann,
  Sturmann, Turner, Farrington, Goldfinger, \& Berger}]{Boyajian2012}
Boyajian, T.~S., von Braun, K., van Belle, G., {et~al.} 2012, The Astrophysical
  Journal, 757, 112

\bibitem[{Boyajian {et~al.}(2013)Boyajian, von Braun, van Belle, Farrington,
  Schaefer, Jones, White, McAlister, ten Brummelaar, Ridgway, Gies, Sturmann,
  Sturmann, Turner, Goldfinger, \& Vargas}]{Boyajian2013}
---. 2013, The Astrophysical Journal, 771, 40

\bibitem[{Brown {et~al.}(2011)Brown, Latham, Everett, \& Esquerdo}]{Brown2011}
Brown, T.~M., Latham, D.~W., Everett, M.~E., \& Esquerdo, G.~A. 2011, The
  Astronomical Journal, 142, 112

\bibitem[{Crossfield {et~al.}(2015)Crossfield, Petigura, Schlieder, Howard,
  Fulton, Aller, Ciardi, L{\'{e}}pine, Barclay, de~Pater, de~Kleer, Quintana,
  Christiansen, Schlafly, Kaltenegger, Crepp, Henning, Obermeier, Deacon,
  Weiss, Isaacson, Hansen, Liu, Greene, Howell, Barman, \&
  Mordasini}]{Crossfield2015}
Crossfield, I. J.~M., Petigura, E., Schlieder, J.~E., {et~al.} 2015, The
  Astrophysical Journal, 804, 10

\bibitem[{Cutri {et~al.}(2006)Cutri, Skrutskie, {Van Dyk}, Beichman, Carpenter,
  Chester, Cambresy, Evans, Fowler, Gizis, Howard, Huchra, Jarrett, Kopan,
  Kirkpatrick, Light, Marsh, McCallon, Schneider, Stiening, Sykes, Weinberg,
  Wheaton, Wheelock, \& Zacharias}]{Cutri2006}
Cutri, R., Skrutskie, M., {Van Dyk}, S., {et~al.} 2006, {Explanatory Supplement
  to the 2MASS All Sky Data Release}

\bibitem[{Cutri {et~al.}(2003)Cutri, Skrutskie, van Dyk, Beichman, Carpenter,
  Chester, Cambresy, Evans, Fowler, Gizis, Howard, Huchra, Jarrett, Kopan,
  Kirkpatrick, Light, Marsh, McCallon, Schneider, Stiening, Sykes, Weinberg,
  Wheaton, Wheelock, \& Zacarias}]{Cutri2003}
Cutri, R.~M., Skrutskie, M.~F., van Dyk, S., {et~al.} 2003, VizieR Online Data
  Catalog, 2246

\bibitem[{Delfosse {et~al.}(2000)Delfosse, Forveille, S{\'{e}}gransan, Beuzit,
  Udry, Perrier, \& Mayor}]{Delfosse2000}
Delfosse, X., Forveille, T., S{\'{e}}gransan, D., {et~al.} 2000, Astronomy and
  Astrophysics, 364, 217

\bibitem[{Delfosse {et~al.}(2013)Delfosse, Bonfils, Forveille, Udry, Mayor,
  Bouchy, Gillon, Lovis, Neves, Pepe, Perrier, Queloz, Santos, \&
  S{\'{e}}gransan}]{Delfosse2013}
Delfosse, X., Bonfils, X., Forveille, T., {et~al.} 2013, Astronomy {\&}
  Astrophysics, 553, A8

\bibitem[{Dittmann {et~al.}(2014)Dittmann, Irwin, Charbonneau, \&
  Berta-Thompson}]{Dittmann2014}
Dittmann, J.~A., Irwin, J.~M., Charbonneau, D., \& Berta-Thompson, Z.~K. 2014,
  The Astrophysical Journal, 784, 156

\bibitem[{Dittmann {et~al.}(2016)Dittmann, Irwin, Charbonneau, \&
  Newton}]{Dittmann2015}
Dittmann, J.~A., Irwin, J.~M., Charbonneau, D., \& Newton, E.~R. 2016, 
  The Astrophysical Journal, 818, 153

\bibitem[{Dressing \& Charbonneau(2013)}]{Dressing2013}
Dressing, C.~D., \& Charbonneau, D. 2013, The Astrophysical Journal, 767, 95

\bibitem[{Dumusque {et~al.}(2011)Dumusque, Lovis, S{\'{e}}gransan, Mayor, Udry,
  Benz, Bouchy, {Lo Curto}, Mordasini, Pepe, Queloz, Santos, \&
  Naef}]{Dumusque2011}
Dumusque, X., Lovis, C., S{\'{e}}gransan, D., {et~al.} 2011, Astronomy {\&}
  Astrophysics, 535, A55

\bibitem[{Dumusque {et~al.}(2012)Dumusque, Pepe, Lovis, S{\'{e}}gransan,
  Sahlmann, Benz, Bouchy, Mayor, Queloz, Santos, \& Udry}]{Dumusque2012}
Dumusque, X., Pepe, F., Lovis, C., {et~al.} 2012, Nature, 491, 207

\bibitem[{Feroz \& Hobson(2013)}]{Feroz2013}
Feroz, F., \& Hobson, M.~P. 2013, Monthly Notices of the Royal Astronomical
  Society, 437, 3540

\bibitem[{Forveille {et~al.}(2011)Forveille, Bonfils, Delfosse, Alonso, Udry,
  Bouchy, Gillon, Lovis, Neves, Mayor, Pepe, Queloz, Santos, Segransan,
  Almenara, Deeg, \& Rabus}]{Forveille2011}
Forveille, T., Bonfils, X., Delfosse, X., {et~al.} 2011, eprint arXiv:1109.2505

\bibitem[{Gaidos {et~al.}(2014)Gaidos, Mann, Lepine, Buccino, James, Ansdell,
  Petrucci, Mauas, \& Hilton}]{Gaidos2014}
Gaidos, E., Mann, A.~W., Lepine, S., {et~al.} 2014, Monthly Notices of the
  Royal Astronomical Society, 443, 2561

\bibitem[{Gray(2009)}]{Gray2009}
Gray, D.~F. 2009, The Astrophysical Journal, 697, 1032

\bibitem[{Gregory(2011)}]{Gregory2011}
Gregory, P.~C. 2011, Monthly Notices of the Royal Astronomical Society, 415,
  2523

\bibitem[{Hartman {et~al.}(2011)Hartman, Bakos, Noyes, Sipőcz, Kov{\'{a}}cs,
  Mazeh, Shporer, \& P{\'{a}}l}]{Hartman2011}
Hartman, J.~D., Bakos, G.~{\'{A}}., Noyes, R.~W., {et~al.} 2011, The
  Astronomical Journal, 141, 166

\bibitem[{Hatzes(2013)}]{Hatzes2013}
Hatzes, A.~P. 2013, Astronomische Nachrichten, 334, 616

\bibitem[{Henry {et~al.}(1997)Henry, Baliunas, Donahue, Soon, \&
  Saar}]{Henry1997}
Henry, G.~W., Baliunas, S.~L., Donahue, R.~A., Soon, W.~H., \& Saar, S.~H.
  1997, The Astrophysical Journal, 474, 503

\bibitem[{Irwin {et~al.}(2014)Irwin, Berta-Thompson, Charbonneau, Dittmann,
  Falco, Newton, \& Nutzman}]{Irwin2014}
Irwin, J.~M., Berta-Thompson, Z.~K., Charbonneau, D., {et~al.} 2014, 18th
  Cambridge Workshop on Cool Stars, Stellar Systems, and the Sun, Proceedings
  of the conference held at Lowell Observatory, 8-14 June, 2014. Edited by G.
  van Belle and H.C. Harris., 767

\bibitem[{Jao {et~al.}(2005)Jao, Henry, Subasavage, Brown, Ianna, Bartlett,
  Costa, \& M{\'{e}}ndez}]{Jao2005}
Jao, W.-C., Henry, T.~J., Subasavage, J.~P., {et~al.} 2005, The Astronomical
  Journal, 129, 1954

\bibitem[{Kiraga \& Stepien(2007)}]{Kiraga2007}
Kiraga, M., \& Stepien, K. 2007, Acta Astronomica, 57, 149

\bibitem[{Kopparapu {et~al.}(2013)Kopparapu, Ramirez, Kasting, Eymet, Robinson,
  Mahadevan, Terrien, Domagal-Goldman, Meadows, \& Deshpande}]{Kopparapu2013}
Kopparapu, R.~K., Ramirez, R., Kasting, J.~F., {et~al.} 2013, The Astrophysical
  Journal, 765, 131

\bibitem[{Leggett(1992)}]{Leggett1992}
Leggett, S.~K. 1992, The Astrophysical Journal Supplement Series, 82, 351

\bibitem[{L{\'{e}}pine {et~al.}(2013)L{\'{e}}pine, Hilton, Mann, Wilde,
  Rojas-Ayala, Cruz, \& Gaidos}]{Lepine2013}
L{\'{e}}pine, S., Hilton, E.~J., Mann, A.~W., {et~al.} 2013, The Astronomical
  Journal, 145, 102

\bibitem[{L{\'{e}}pine \& Shara(2005)}]{Lepine2005a}
L{\'{e}}pine, S., \& Shara, M.~M. 2005, The Astronomical Journal, 129, 1483

\bibitem[{Mahadevan {et~al.}(2012)Mahadevan, Ramsey, Bender, Terrien, Wright,
  Halverson, Hearty, Nelson, Burton, Redman, Osterman, Diddams, Kasting, Endl,
  \& Deshpande}]{Mahadevan2012}
Mahadevan, S., Ramsey, L., Bender, C., {et~al.} 2012, in Ground-based and
  Airborne Instrumentation for Astronomy IV, ed. I.~S. McLean, S.~K. Ramsay, \&
  H.~Takami, Vol. 8446, 84461S

\bibitem[{Makarov \& Berghea(2014)}]{Makarov2014}
Makarov, V.~V., \& Berghea, C. 2014, The Astrophysical Journal, 780, 124

\bibitem[{Mayor {et~al.}(2009)Mayor, Bonfils, Forveille, Delfosse, Udry,
  Bertaux, Beust, Bouchy, Lovis, Pepe, Perrier, Queloz, \& Santos}]{Mayor2009}
Mayor, M., Bonfils, X., Forveille, T., {et~al.} 2009, Astronomy and
  Astrophysics, 507, 487

\bibitem[{McQuillan {et~al.}(2013)McQuillan, Aigrain, \& Mazeh}]{McQuillan2013}
McQuillan, A., Aigrain, S., \& Mazeh, T. 2013, Monthly Notices of the Royal
  Astronomical Society, 432, 1203

\bibitem[{McQuillan {et~al.}(2014)McQuillan, Mazeh, \& Aigrain}]{McQuillan2014}
McQuillan, A., Mazeh, T., \& Aigrain, S. 2014, The Astrophysical Journal
  Supplement Series, 211, 24

\bibitem[{Meunier {et~al.}(2010)Meunier, Desort, \& Lagrange}]{Meunier2010}
Meunier, N., Desort, M., \& Lagrange, A.-M. 2010, Astronomy and Astrophysics,
  512, A39

\bibitem[{Mohanty \& Basri(2003)}]{Mohanty2003}
Mohanty, S., \& Basri, G. 2003, The Astrophysical Journal, 583, 451

\bibitem[{Narita {et~al.}(2013)Narita, Fukui, Ikoma, Hori, Kurosaki, Kawashima,
  Nagayama, Onitsuka, Sukom, Nakajima, Tamura, Kuroda, Yanagisawa, Hirano,
  Kawauchi, Kuzuhara, Ohnuki, Suenaga, Takahashi, Izumiura, Kawai, \&
  Yoshida}]{Narita2013}
Narita, N., Fukui, A., Ikoma, M., {et~al.} 2013, The Astrophysical Journal,
  773, 144

\bibitem[{Nascimbeni {et~al.}(2015)Nascimbeni, Mallonn, Scandariato, Pagano,
  Piotto, Micela, Messina, Leto, Strassmeier, Bisogni, \&
  Speziali}]{Nascimbeni2015}
Nascimbeni, V., Mallonn, M., Scandariato, G., {et~al.} 2015, Astronomy and Astrophysics, 579, A113

\bibitem[{Newton {et~al.}(2014)Newton, Charbonneau, Irwin, Berta-Thompson,
  Rojas-Ayala, Covey, \& Lloyd}]{Newton2014}
Newton, E.~R., Charbonneau, D., Irwin, J., {et~al.} 2014, The Astronomical
  Journal, 147, 20

\bibitem[{Newton {et~al.}(2015)Newton, Charbonneau, Irwin, \&
  Mann}]{Newton2015}
Newton, E.~R., Charbonneau, D., Irwin, J., \& Mann, A.~W. 2015, The
  Astrophysical Journal, 800, 85

\bibitem[{Newton {et~al.}(2016)Newton, Irwin, Charbonneau, Berta-Thompson,
  Dittmann, \& West}]{Newton2016}
Newton, E.~R., Irwin, J., Charbonneau, D., {et~al.} 2016, arXiv:1511.00957

\bibitem[{Nutzman \& Charbonneau(2008)}]{Nutzman2008}
Nutzman, P., \& Charbonneau, D. 2008, Publications of the Astronomical Society
  of the Pacific, 120, 317

\bibitem[{Queloz {et~al.}(2009)Queloz, Bouchy, Moutou, Hatzes, H{\'{e}}brard,
  Alonso, Auvergne, Baglin, Barbieri, Barge, Benz, Bord{\'{e}}, Deeg, Deleuil,
  Dvorak, Erikson, {Ferraz Mello}, Fridlund, Gandolfi, Gillon, Guenther,
  Guillot, Jorda, Hartmann, Lammer, L{\'{e}}ger, Llebaria, Lovis, Magain,
  Mayor, Mazeh, Ollivier, P{\"{a}}tzold, Pepe, Rauer, Rouan, Schneider,
  Segransan, Udry, \& Wuchterl}]{Queloz2009}
Queloz, D., Bouchy, F., Moutou, C., {et~al.} 2009, Astronomy and Astrophysics,
  506, 303

\bibitem[{Quirrenbach {et~al.}(2014)Quirrenbach, Amado, Caballero, Mundt,
  Reiners, Ribas, Seifert, Abril, Aceituno, Alonso-Floriano, {Ammler-von Eiff},
  {Antona Jim{\'{e}}nez}, Anwand-Heerwart, Azzaro, Bauer, Barrado, Becerril,
  B{\'{e}}jar, Ben{\'{\i}}tez, Berdi{\~{n}}as, C{\'{a}}rdenas, Casal, Claret,
  Colom{\'{e}}, Cort{\'{e}}s-Contreras, Czesla, Doellinger, Dreizler, Feiz,
  Fern{\'{a}}ndez, Galad{\'{\i}}, G{\'{a}}lvez-Ortiz, Garc{\'{\i}}a-Piquer,
  Garc{\'{\i}}a-Vargas, Garrido, Gesa, {G{\'{o}}mez Galera}, {Gonz{\'{a}}lez
  {\'{A}}lvarez}, {Gonz{\'{a}}lez Hern{\'{a}}ndez}, Gr{\"{o}}zinger,
  Gu{\`{a}}rdia, Guenther, de~Guindos, Guti{\'{e}}rrez-Soto, Hagen, Hatzes,
  Hauschildt, Helmling, Henning, Hermann, {Hern{\'{a}}ndez Casta{\~{n}}o},
  Herrero, Hidalgo, Holgado, Huber, Huber, Jeffers, Joergens, de~Juan, Kehr,
  Klein, K{\"{u}}rster, Lamert, Lalitha, Laun, Lemke, Lenzen, {L{\'{o}}pez del
  Fresno}, {L{\'{o}}pez Mart{\'{\i}}}, L{\'{o}}pez-Santiago, Mall, Mandel,
  Mart{\'{\i}}n, Mart{\'{\i}}n-Ruiz, Mart{\'{\i}}nez-Rodr{\'{\i}}guez, Marvin,
  Mathar, Mirabet, Montes, {Morales Mu{\~{n}}oz}, Moya, Naranjo, Ofir, Oreiro,
  Pall{\'{e}}, Panduro, Passegger, P{\'{e}}rez-Calpena, {P{\'{e}}rez
  Medialdea}, Perger, Pluto, Ram{\'{o}}n, Rebolo, Redondo, Reffert, Reinhardt,
  Rhode, Rix, Rodler, Rodr{\'{\i}}guez, Rodr{\'{\i}}guez-L{\'{o}}pez,
  Rodr{\'{\i}}guez-P{\'{e}}rez, Rohloff, Rosich, S{\'{a}}nchez-Blanco,
  {S{\'{a}}nchez Carrasco}, Sanz-Forcada, Sarmiento, Sch{\"{a}}fer, Schiller,
  Schmidt, Schmitt, Solano, Stahl, Storz, St{\"{u}}rmer, Su{\'{a}}rez, Ulbrich,
  Veredas, Wagner, Winkler, {Zapatero Osorio}, Zechmeister, {Abell{\'{a}}n de
  Paco}, Anglada-Escud{\'{e}}, del Burgo, Klutsch, Lizon, L{\'{o}}pez-Morales,
  Morales, Perryman, Tulloch, \& Xu}]{Quirrenbach2014}
Quirrenbach, A., Amado, P.~J., Caballero, J.~A., {et~al.} 2014, in Proceedings
  of the SPIE, ed. S.~K. Ramsay, I.~S. McLean, \& H.~Takami, Vol. 9147, 91471F

\bibitem[{Rajpaul {et~al.}(2015)Rajpaul, Aigrain, Osborne, Reece, \&
  Roberts}]{Rajpaul2015}
Rajpaul, V., Aigrain, S., Osborne, M.~A., Reece, S., \& Roberts, S. 2015,
  Monthly Notices of the Royal Astronomical Society, 452, 2269

\bibitem[{Riedel {et~al.}(2010)Riedel, Subasavage, Finch, Jao, Henry, Winters,
  Brown, Ianna, Costa, \& Mendez}]{Riedel2010}
Riedel, A.~R., Subasavage, J.~P., Finch, C.~T., {et~al.} 2010, The Astronomical
  Journal, 140, 897

\bibitem[{Rivera {et~al.}(2005)Rivera, Lissauer, Butler, Marcy, Vogt, Fischer,
  Brown, Laughlin, \& Henry}]{Rivera2005}
Rivera, E.~J., Lissauer, J.~J., Butler, R.~P., {et~al.} 2005, The Astrophysical
  Journal, 634, 625

\bibitem[{Robertson {et~al.}(2015{\natexlab{a}})Robertson, Endl, Henry,
  Cochran, MacQueen, \& Williamson}]{Robertson2015a}
Robertson, P., Endl, M., Henry, G.~W., {et~al.} 2015{\natexlab{a}}, The
  Astrophysical Journal, 801, 79

\bibitem[{Robertson \& Mahadevan(2014)}]{Robertson2014a}
Robertson, P., \& Mahadevan, S. 2014, The Astrophysical Journal, 793, L24

\bibitem[{Robertson {et~al.}(2014)Robertson, Mahadevan, Endl, \&
  Roy}]{Robertson2014}
Robertson, P., Mahadevan, S., Endl, M., \& Roy, A. 2014, Science (New York,
  N.Y.), 345, 440

\bibitem[{Robertson {et~al.}(2015{\natexlab{b}})Robertson, Roy, \&
  Mahadevan}]{Robertson2015}
Robertson, P., Roy, A., \& Mahadevan, S. 2015{\natexlab{b}}, The Astrophysical
  Journal, 805, L22

\bibitem[{Rodler \& L{\'{o}}pez-Morales(2014)}]{Rodler2014}
Rodler, F., \& L{\'{o}}pez-Morales, M. 2014, The Astrophysical Journal, 781, 54

\bibitem[{Saar \& Donahue(1997)}]{Saar1997}
Saar, S.~H., \& Donahue, R.~A. 1997, The Astrophysical Journal, 485, 319

\bibitem[{Schneider {et~al.}(2011)Schneider, Dedieu, {Le Sidaner}, Savalle, \&
  Zolotukhin}]{Schneider2011}
Schneider, J., Dedieu, C., {Le Sidaner}, P., Savalle, R., \& Zolotukhin, I.
  2011, Astronomy {\&} Astrophysics, 532, A79

\bibitem[{Skumanich(1972)}]{Skumanich1972}
Skumanich, A. 1972, The Astrophysical Journal, 171, 565

\bibitem[{Snellen {et~al.}(2013)Snellen, de~Kok, le~Poole, Brogi, \&
  Birkby}]{Snellen2013}
Snellen, I. A.~G., de~Kok, R.~J., le~Poole, R., Brogi, M., \& Birkby, J. 2013,
  The Astrophysical Journal, 764, 182

\bibitem[{{Su{\'{a}}rez Mascare{\~{n}}o} {et~al.}(2015){Su{\'{a}}rez
  Mascare{\~{n}}o}, Rebolo, {Gonz{\'{a}}lez Hern{\'{a}}ndez}, \&
  Esposito}]{SuarezMascareno2015}
{Su{\'{a}}rez Mascare{\~{n}}o}, A., Rebolo, R., {Gonz{\'{a}}lez
  Hern{\'{a}}ndez}, J.~I., \& Esposito, M. 2015, Monthly Notices of the Royal
  Astronomical Society, 452, 2745

\bibitem[{Terrien {et~al.}(2015)Terrien, Mahadevan, Deshpande, \&
  Bender}]{Terrien2015}
Terrien, R.~C., Mahadevan, S., Deshpande, R., \& Bender, C.~F. 2015, The
  Astrophysical Journal Supplement Series, 220, 16

\bibitem[{Tuomi(2011)}]{Tuomi2011}
Tuomi, M. 2011, Astronomy {\&} Astrophysics, 528, L5

\bibitem[{Udry {et~al.}(2007)Udry, Bonfils, Delfosse, Forveille, Mayor,
  Perrier, Bouchy, Lovis, Pepe, Queloz, \& Bertaux}]{Udry2007}
Udry, S., Bonfils, X., Delfosse, X., {et~al.} 2007, Astronomy and Astrophysics,
  469, L43

\bibitem[{van Leeuwen(2007)}]{VanLeeuwen2007}
van Leeuwen, F. 2007, Astronomy and Astrophysics, 474, 653

\bibitem[{Vogt {et~al.}(2010)Vogt, Butler, Rivera, Haghighipour, Henry, \&
  Williamson}]{Vogt2010}
Vogt, S.~S., Butler, R.~P., Rivera, E.~J., {et~al.} 2010, The Astrophysical
  Journal, 723, 954

\bibitem[{West {et~al.}(2008)West, Hawley, Bochanski, Covey, Reid, Dhital,
  Hilton, \& Masuda}]{West2008}
West, A.~A., Hawley, S.~L., Bochanski, J.~J., {et~al.} 2008, The Astronomical
  Journal, 135, 785

\bibitem[{West {et~al.}(2015)West, Weisenburger, Irwin, Berta-Thompson,
  Charbonneau, Dittmann, \& Pineda}]{West2015}
West, A.~A., Weisenburger, K.~L., Irwin, J., {et~al.} 2015, The Astrophysical
  Journal, 812, 3

\end{thebibliography}
\end{document}